\documentclass[a4paper,aps,prl,showpacs,showkeywords,twocolumn]{revtex4}

\usepackage{amssymb}
\usepackage{graphicx}
\usepackage{amsmath}
\usepackage{amsbsy}
\usepackage{amsthm}
\usepackage{bbm}
\usepackage{bm}
\usepackage{epsfig}

\newcommand{\be}{\begin{equation}}
\newcommand{\ee}{\end{equation}}
\newcommand{\beq}{\begin{eqnarray}}
\newcommand{\eeq}{\end{eqnarray}}
\newcommand{\bra}[1]{\ensuremath{\langle #1 |}}
\newcommand{\ket}[1]{\ensuremath{| #1 \rangle}}
\usepackage{geometry}
\begin{document}

\title{Proving the Generation of Genuine Multipartite Entanglement in a Single-Neutron Interferometer Experiment}

\author{Daniel Erd\"{o}si$^{1}$}
\email{derdosi@ati.ac.at}
\author{Marcus Huber$^{2,3}$}
\email{marcus.huber@univie.ac.at}
\author{Beatrix C. Hiesmayr$^{2,4}$}
\email{beatrix.hiesmayr@univie.ac.at}
\author{Yuji Hasegawa$^{1}$}
\email{hasegawa@ati.ac.at}
\affiliation{$^{1}$Atominstitut, Vienna University of Technology, Stadionallee 2, A-1020 Vienna, Austria}
\affiliation{$^{2}$University of Vienna, Faculty of Physics, Boltzmanngasse 5, 1090 Vienna, Austria}
\affiliation{$^{3}$University of Bristol, Department of Mathematics, Bristol, BS8 1TW, U.K.}
\affiliation{$^{4}$Masaryk University, Department of Theoretical Physics and Astrophysics, Kotl\'a\v{r}\'ska 2, 61137 Brno, Czech Republic}

\begin{abstract}
We present experimental evidence of the generation of distinct types of genuine multipartite entanglement between three degrees of freedom (spin, energy, and path) within single-neutron quantum systems. This is achieved via the development of new spin manipulation apparatuses for neutron interferometry and the entanglement is detected via appropriately designed and optimized non-linear witnesses. Via the applied criteria we reveal even finer properties of the type of the genuine multipartite entanglement that is produced in the experiment. In a subsequent analysis we show the extraordinarily high fidelity of the generated entangled states, proving the outstanding level of controllability of these systems.
\end{abstract}


\maketitle

\section{Introduction}

Entanglement is at the heart of the quantum information and communication technology \cite{NC}. It has been found in many distinct physical systems, e.g., photons, atoms, ions, neutrons, and even in quark-antiquark systems of the second and third generation (e.g.~\cite{Hiesmayr1}). The simplest entanglement is achieved with a bipartite system, while multipartite entanglement has been shown to enable different applications, e.g., secure communications between many parties (e.g.~\cite{HBB,SHH1,gisin-crypt}) or the basis for measurement based quantum computation (first introduced in \cite{qc}). In addition, it may play a fundamental role in various complex physical systems ranging from condensed matter systems (e.g.~\cite{phase,helium,hiesmayrnarnhofer,illuminati}) to possibly even biological ones (e.g.~\cite{bio3}). Here, it is worth noting that entanglement of multipartite systems is essentially different from entanglement of bipartite systems. Multipartite entanglement, with which this paper deals, can e.g. be used for the study of quantum contextuality \cite{contextuality}, whereas entanglement between different particles (multiparticle entanglement) can also be used to study quantum non-locality.

In order to explore further sophisticated applications it is crucial to have a well founded understanding of multipartite entanglement as well as to find systems in which a controllable physical implementation is feasible. A single-neutron system provides an extraordinarily controllable three qubit system and as such is perfectly suited to study the structure of tripartite Hilbert spaces and all predictions for correlations therein. We demonstrate multipartite entanglement by means of this system and in principle open up a path to high precision tests of three qubit Hilbert spaces. For known physical applications that require genuine multipartite entanglement, as e.g. quantum computing, intra-particle entanglement is probably not the first choice, however, for purposes of fundamental tests, which we aim for in this paper, intra-particle entanglement serves just as well. In terms of potential physical applications most progress has been made in photonic systems, as the techniques for manipulating photons locally is very well developed \cite{Photon1}. Photons move at the speed of light, making them ideally suited for long range communication (such as quantum secret sharing). Nevertheless, although some attempts at photonic quantum computer have successfully been made, it is not trivial to hold and store photons at a place, which is an essential requirement for quantum computing of higher classes. In that respect other approaches have been pursued ranging from ion traps \cite{ions} to proposals for quantum optomechanical systems \cite{optom}.

Most physical implementations of multipartite entangled systems explore only one kind of internal degree of freedom that is entangled to the same internal degree of freedom of other physically disjoint systems at different locations. Another kind of entanglement is reported by entangling photons in all possible degrees of freedom, which is called hyper-entanglement \cite{Martini}. Entanglement purely between degrees of freedom in a single quantum system is extensively studied by using neutron optical techniques in particular neutron interferometry \cite{Rauchbook}: starting with a demonstration of the violation of a Bell-like inequality \cite{HaseN}, the implementation of a triply entangled Greenberger-Zeilinger-Horne (GHZ)-like state is reported \cite{HasePRA}. Quite recently, a test of an alternative model of quantum mechanics \'a la Leggett \cite{HaseNJP} as well as high-efficiency manipulations of a tripartite-entangled system \cite{Sponar} are reported by using another strategy, namely, neutron polarimetry. It is worth noting here that these neutron optical techniques exploit matter-waves and are very suitable to study properties of multipartite entangled systems of matter-waves: the efficiency of manipulating degrees of freedom in a single neutron is rather high, which enabled tests of alternative models of quantum mechanics with matter-waves.

\section{Theory}

Multipartite entanglement is defined --as in the bipartite case-- via an exclusion of separability. Generally, an $n$-partite pure quantum state is called $k$-separable iff it can be written as a product of $k$ substates, i.e., $|\psi_{k-\mathrm{sep}}\rangle=|\phi_1\rangle\otimes|\phi_2\rangle \otimes\cdots\otimes|\phi_k\rangle$. For mixed states there exists a straightforward definition, namely, a mixed state is $k$-separable iff it has a decomposition into $k$-separable pure states, i.e., $\rho_{k-\mathrm{sep}}=\sum_i p_i |\psi_{k-\mathrm{sep}}^i\rangle\langle\psi_{k-\mathrm{sep}}^i|$. Entangled states consisting of more than two parties can be \emph{partially} separable: it may be that they only are separable with respect to certain partitions, while for other partitions they are entangled. An important class of states pertains to genuinely multipartite entangled states, which are not biseparable (2-separable) with respect to all possible partitions. A biseparable state is one that can be decomposed into a convex set of biseparable pure states. It is worth noting here that the partitions where the decomposed states are biseparable do not need to be the same for each decomposition: e.g., the tripartite state $\rho=p\rho_{1}\otimes\rho_{23}+(1-p)\rho_{12}\otimes\rho_{3}$, with $p\in[0,1]$, is biseparable, though there is no specific bipartition of the whole state (with respect to which the state is separable). Necessary and sufficient criteria to detect biseparability (i.e., entanglement) are only known for the simplest bipartite case. Multipartite entanglement, on the other hand, is much richer in variety and more difficult to detect, since it does not need to be biseparable with respect to a specific decomposition. Various techniques to detect multipartite entanglement have been developed (e.g.~in \cite{crit1,crit6,crit7,crit8}). We adapt a specific technique from \cite{crit6,crit7,crit8}.

Two famous examples of genuine multipartite entanglement, which have been shown to be very different in nature (e.g.~Ref.~\cite{crit1}), are the GHZ-state and the W-state (introduced in Ref.~\cite{GHZstate} and \cite{Wstate}). These are the two states for which we present the experimental proof of production in the tripartite neutron system via multipartite detection criteria. In computational notation they read ($a,b,c,d,\textrm{e}$\dots amplitudes to be determined)
\begin{align}
|W\rangle=a\underbrace{|101\rangle}_{:=|w_1\rangle}+b\;e^{i\pi/2}\underbrace{|011\rangle}_{:=|w_2\rangle}+c\underbrace{|002\rangle}_{:=|w_3\rangle} \label{Wstate}\\
|GHZ\rangle=d\;e^{i\pi/2}|101\rangle+\textrm{e}|010\rangle \label{GHZstate}
\end{align}
where the three subsystems denote three different properties of single neutrons. The first subsystem denotes the path in the interferometer (path degree of freedom), which we also denote as path I ($|{\rm I}\rangle=|0\rangle$) and II ($|{\rm II}\rangle=|1\rangle$), the second denotes the spin of the neutron (which can take the two values $\pm  \frac{\hbar}{2}$ and we denote as $|\downarrow\rangle=|0\rangle$ and $|\uparrow\rangle=|1\rangle$) and the third subsystem refers to the total energy (potential energy in an applied magnetic field plus kinetic energy) of the neutron (where the neutron can be at different total energy levels, denoted as $|E_0\rangle=|0\rangle$, $|E_0-\hbar\omega\rangle=|1\rangle$ and $|E_0-2\hbar\omega\rangle=|2\rangle$).

For the detection of these two states we employ two distinct nonlinear entanglement witnesses, which are optimally suited to detect GHZ- and W-like entanglement. The GHZ witness inequality employs a permutation operator $\Pi_i$, permuting the $i^{th}$ subspace of the two copy Hilbert space, reads
\begin{eqnarray}\label{IGHZ}
\lefteqn{I_{GHZ}:=}\\ \nonumber
&&|\langle010|\rho|101\rangle|-\sum_{i=1}^3\sqrt{\langle010101|\Pi_i\rho^{\otimes 2}\Pi_i|010101\rangle}\leq 0,
\end{eqnarray}
and is satisfied for all biseparable states and maximally violated for the above defined GHZ-state. For the W-state a similar derivation leads to the witness inequality
\begin{eqnarray}\label{IW}
\lefteqn{\lefteqn{I_{W}:=}}\\ \nonumber
&&\sum_{i\neq j}|\langle w_i|\rho|w_j\rangle|-\sum_{i,j=1}^3\sqrt{\langle w_i w_j|\Pi_i\rho^{\otimes 2}\Pi_i|w_i w_j\rangle}\leq 0,
\end{eqnarray}
which is again satisfied for all biseparable states and maximally violated for the above defined W-state [with $|w_i\rangle$ as defined in (\ref{Wstate})]. Thus any violation of any of these two inequalities is a clear proof of the presence of genuine multipartite entanglement within the system. Furthermore both witnesses were recently shown to provide a good lower bound for an entropic measure of genuine multipartite entanglement~\cite{crazychin,wuetal}. To detect $k$-separability, we employ the following inequality, which is satisfied by all $k$-separable states \cite{crit7}:
\begin{eqnarray}\label{k-sep}
I_{k-sep}[\rho]:&=& \sqrt{\bra{\Phi}\rho^{\otimes 2} P_{total}\ket{\Phi}} \\
&& - \sum_{\{\alpha\}}\left(\prod_{i=1}^k \bra{\Phi}P_{\alpha_i}^\dagger \rho^{\otimes 2} P_{\alpha_i} \ket{\Phi}\right)^{\frac{1}{2k}} \leq 0, \nonumber
\end{eqnarray}
where $\ket{\Phi}=\ket{\phi_1}\otimes\ket{\phi_2}$ is an arbitrary fully separable state, $P_{\alpha_i}$ is a permutation operator permuting the $\alpha_i$-th elements of $\ket{\phi_1}$ and $\ket{\phi_2}$, $P_{total}$ wholly permutes the $\ket{\phi_i}$, and the sum runs over all $k$-partitions $\{\alpha\}$.

\section{Experiment}

\subsection{Preparation of W- and Greenberger-Zeilinger-Horne-like states}

The experiment, illustrated in Fig.~\ref{setup}, was carried out at the perfect-crystal neutron optics beam line S18 at the high flux reactor of the Institute Laue-Langevin (ILL) \cite{KROUPA}. By means of a Si perfect-crystal monochromator, a neutron beam with a mean wavelength of $\lambda_{0}=1.92$\AA\ ($\Delta\lambda / \lambda_0 \approx 0.02$) was selected. Then the beam was polarized vertically by magnetic-prism refractions. Because of the angular separation of the two sub-beams, only up-spin neutrons $|\uparrow\rangle$ met the Bragg condition at the first interferometer plate. After this the beam, with an aperture of 4x8mm$^2$, entered a skew-symmetric silicon perfect crystal neutron interferometer (IFM), which was adjusted to give a 220 reflection, thus coherently splitting the incident beam into the two spatially separated paths I and II at the first IFM plate. A parallel-sided Al plate was used as a phase shifter to vary the relative phase $\chi$ for the path degree of freedom prior to the coherent recombination of the two paths at the last IFM plate. A pair of water-cooled Helmholtz coils produced a fairly uniform magnetic guide field $B_{0}\hat{\bf z}$ of 2.3mT over the region of the polarized neutron beam. This field also defined a potential energy (Zeeman energy) of the neutrons, while the total energy of a neutron is equal to the sum of its potential energy and kinetic energy. Our energy degree of freedom is the total energy, as already utilized in our previous experiment \cite{Sponar08a}. To manipulate the neutrons, we used radio-frequency (RF) spin flippers, which change both the spin and the total energy, and a direct current (DC) spin flipper, which changes the spin without changing the total energy. The change in total energy involved by a flip from spin-up to spin-down due to an RF coil operating at frequency $\omega$ is given by $-\hbar\omega$. By a $\pi$ flip we mean a flip at that most neutrons' spins are flipped (according to the efficiency of the spin flipping device), whereas for a $\pi${\small /2} flip there is only a 50\% probability for a neutron to be subject to a spin flip. Each time a neutron suffers a spin flip, it also suffers a phase shift $e^{i\pi/2}$. In path I of the IFM two RF spin flippers were placed in a row, the first one RF$_{\pi}^{2\omega}$ operating at double frequency $2\omega$ and performing a $\pi$ flip, the second one RF$_{\pi/2}^{\omega}$ operating at the frequency $\omega$ and performing a $\pi${\small /2} flip, whereas in path II one RF flipper RF$_{\pi}^{\omega}$ was placed, operating at the frequency $\omega$ and performing a $\pi$ flip (see Fig.~\ref{setup}). RF$_{\pi}^{2\omega}$ was equipped with its own Helmholtz coil pair, which enabled it to operate at double frequency. Due to a new coil cooling system, thermal disturbances from the coils in the IFM were negligible.

\begin{widetext}
\begin{center}
\begin{figure}[htb]
\resizebox{0.8\columnwidth}{!}{\includegraphics{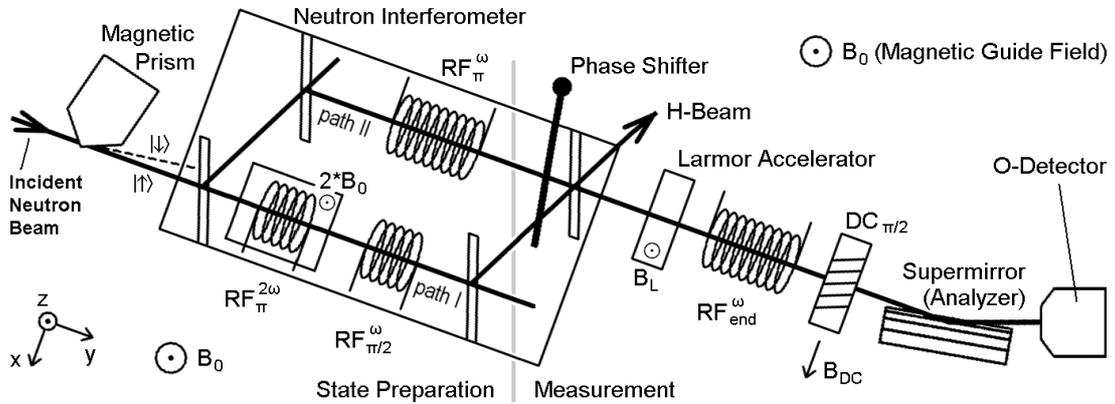}}
\caption{Experimental setup (see details in text).}
\label{setup}
\end{figure}
\end{center}
\end{widetext}

For the W- and the GHZ-like state we entangled the path, spin, and energy degrees of freedom of single neutrons. The state was generated as follows: The spin-up polarized neutron beam $\left|\uparrow 0\right\rangle$ entered the IFM and split into two partial beams $\frac{1}{\sqrt{2}}\left|{\rm I}\uparrow 0\right\rangle$ and $\frac{1}{\sqrt{2}}\left|{\rm II}\uparrow 0\right\rangle$. In the ideal case, RF$_{\pi}^{2\omega}$ changes $\frac{1}{\sqrt{2}}\left|{\rm I}\uparrow 0\right\rangle$ to $\frac{1}{\sqrt{2}}e^{i\pi/2}\left|{\rm I}\downarrow -2\hbar\omega\right\rangle$, which then the subsequent RF$_{\omega}^{\pi/2}$ changes to $\frac{1}{\sqrt{4}}e^{i\pi/2}\left|{\rm I}\downarrow -2\hbar\omega\right\rangle + \frac{1}{\sqrt{4}}e^{i\pi}\left|{\rm I}\uparrow -\hbar\omega\right\rangle$, while RF$_{\pi}^{\omega}$ changes $\frac{1}{\sqrt{2}}\left|{\rm II}\uparrow 0\right\rangle$ to $\frac{1}{\sqrt{2}}e^{i\pi/2}\left|{\rm II}\downarrow -\hbar\omega\right\rangle$. For the symmetric (i.e., balanced) W-state ($|W_{sym}\rangle$), with theoretical nominal amplitudes $a=b=c=\frac{1}{\sqrt{3}}$, a 50\% stochastic absorber (an indium plate with thickness $0.75$mm) was placed in path II, whereas without the absorber an asymmetric (i.e., unbalanced) W-like state ($|W_{asym}\rangle$) was generated with theoretical nominal amplitudes $a = \frac{1}{\sqrt{2}}$ and $b = c = \frac{1}{\sqrt{4}}$. For the preparation of the GHZ-like state we expect theoretical nominal amplitudes $d = \textrm{e} = \frac{1}{\sqrt{2}}$. Here, only RF$_{\pi}^{\omega}$ was switched on for state preparation while RF$_{\pi}^{2\omega}$ and RF$_{\pi/2}^{\omega}$ did not operate.

\subsection{Measurement}

Only the O-beam was used for the measurements, where a spin analyzing super-mirror (transmitting up-spin neutrons only) together with a Larmor accelerator, an RF flipper RF$_\mathrm{end}^\omega$, operated at the frequency $\omega$, and a DC flipper DC$_{\pi/2}$, performing a $\pi${\small /2} flip, enabled the selection of neutrons in dependence of their polarization and their energy for detection. Neutrons with the selected spin properties were counted in the subsequent O-detector.

A full tomography of the generated states would have been too complex and partly even infeasible (because certain coherences cannot be achieved). As entanglement witnesses do not require a full state tomography they clearly provide an advantage in this case.

For the W-state, Eq.~(\ref{Wstate}), the terms $\langle101|\rho|101\rangle$, $\langle011|\rho|011\rangle$, $\langle002|\rho|002\rangle$, $|\langle011|\rho|101\rangle|$, $|\langle002|\rho|101\rangle|$, and $|\langle002|\rho|011\rangle|$ were determined. The terms $\langle101|\rho|101\rangle$, $\langle011|\rho|011\rangle$, and $\langle002|\rho|002\rangle$ were determined by intensity measurements while blocking one path in the IFM (the Larmor accelerator and DC$_{\pi/2}$ did not operate). For the measurement of $\langle101|\rho|101\rangle$, path I of the IFM was blocked while in path II the beam was manipulated by RF$_{\pi}^{\omega}$, and the analysis setup only consisted of the super-mirror. The reference measurement was obtained while RF$_{\pi}^{\omega}$ was switched off. For the measurement of $\langle011|\rho|011\rangle$ and $\langle002|\rho|002\rangle$, path II of the IFM was blocked while in path I the beam was manipulated by RF$_{\pi}^{2\omega}$ and RF$_{\pi/2}^{\omega}$. Spin-up, spin-down and a reference measurement were made. For the spin-down measurement, RF$_\mathrm{end}^\omega$ performed a $\pi$ flip, while at the two other measurements it was switched off and for the reference measurement also RF$_{\pi}^{2\omega}$ and RF$_{\pi/2}^{\omega}$ were switched off. The mathematical relations between the measured intensities and $\langle101|\rho|101\rangle$, $\langle011|\rho|011\rangle$, and $\langle002|\rho|002\rangle$ can be derived from (\ref{Wstate}) by considering the manipulations of the neutron beam in the analysis setup by the beam blocker, RF$_\mathrm{end}^\omega$, and the super-mirror (here we had to assume that all devices manipulating the beam line were operating as previously determined). Each measurement was carried out $10$ times in order to reduce statistical errors.

\begin{widetext}
\begin{center}
\begin{figure}[htb]
\resizebox{0.8\columnwidth}{!}{\includegraphics{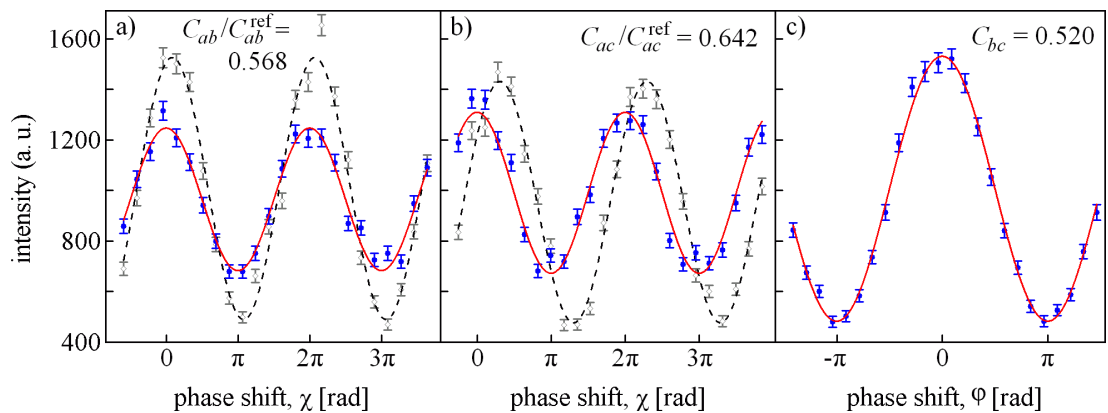}}
\caption{Typical sinusoidal oscillations of the count rates: a) and b) show contrast scans for $|\langle011|\rho|101\rangle|$ and $|\langle002|\rho|101\rangle|$, respectively, together with the reference oscillations (dashed curves) and c) illustrates a contrast scan for $|\langle002|\rho|011\rangle|$.}
\label{example}
\end{figure}
\end{center}
\end{widetext}

In the experimental setups for the determination of the cross terms $|\langle011|\rho|101\rangle|$, $|\langle002|\rho|101\rangle|$, and $|\langle002|\rho|011\rangle|$ both paths of the IFM were open and we determined these terms by means of the contrasts $C$ of three sinusoidal oscillations while varying the relative phase between coherent terms. Coherence between each two of the three terms of (\ref{Wstate}) was achieved by three different analysis setups. The determination of $|\langle011|\rho|101\rangle|$ and $|\langle002|\rho|101\rangle|$ requires the coherence of terms that are generated in different paths of the IFM, therefore contrast scans were made by successively turning the phase shifter to successively change the relative phase $\chi$ between the respective two terms. The two terms for $|\langle011|\rho|101\rangle|$ have the same energy but different spins, therefore in the analysis setup DC$_{\pi/2}$ was switched on, so that the two resulting spin-up terms could interfere \cite{Sponar08a}. For the measurement of $|\langle002|\rho|101\rangle|$ two terms have to interfere that also have an energy difference of $\hbar\omega$, therefore also RF$_\mathrm{end}^\omega$ performed a $\pi${\small /2} flip, which transformed the two terms in question into energetically equal and after DC$_{\pi/2}$ into coherent ones. At each point of these contrast measurements also a reference point was recorded, where the three RF flippers in the IFM were switched off, from which reference contrasts $C^\mathrm{ref}$ were obtained (the average of the reference contrasts amounts to $0.455$). The contrasts were divided by the corresponding reference contrasts in the measurement evaluation. $|\langle002|\rho|011\rangle|$, whose components were generated in the same IFM path and had different energies, was determined by changing the components' relative phase $\phi$ by successively changing the strength of the magnetic field in the Larmor accelerator and then making the components coherent by RF$_\mathrm{end}^\omega$ performing a $\pi/2$ flip. The mathematical relations between the measured contrasts and the three cross terms can be derived from (\ref{Wstate}) by considering the manipulations of the neutron beam in the analysis setup: $2|\langle011|\rho|101\rangle|=C_{ab}/C_{ab}^\mathrm{ref}$, $|\langle002|\rho|101\rangle|=C_{ac}/C_{ac}^\mathrm{ref}-|\langle011|\rho|101\rangle|$, and $2|\langle002|\rho|011\rangle|=C_{bc}$. Each measurement was carried out 4 times in order to reduce statistical errors and the contrasts and their errors were determined from least squares fits. Typical contrast scans are shown in Fig.~\ref{example}.

We carried out these measurements without and with the 50\% stochastic absorber in path II, yielding the asymmetric and the symmetric W-like state, respectively. In the first case, each point was measured for 60s, whereas in the latter case for 80s, to compensate for the total intensity loss due to the absorber. To test the robustness of the generated states, all these measurements were repeated under degrading of coherence between the beams in path I and II. The loss of coherence was achieved by a shift of the wave packets relative to each other due to an aluminium plate placed in path II. For the GHZ-like state, (\ref{GHZstate}), the terms $\langle101|\rho|101\rangle$, $\langle010|\rho|010\rangle$, and $|\langle101|\rho|010\rangle|$ were measured. The determinations of these terms are analogous to those described above.

\section{Results: determination of entanglement witness}

The fidelities achieved for the GHZ-like state and the symmetric and asymmetric W-like states were $0.985\pm 0.011$, $0.987\pm 0.029$, and $0.948\pm 0.022$, respectively, which values can be mainly attributed to the imperfect efficiencies of the spin flippers. The violation of the inequalities (\ref{IGHZ}) and (\ref{IW}) listed in Tab.~\ref{TableI} clearly demonstrates the presence of genuine multipartite entanglement. The fidelities achieved for the respective degraded W-states were $0.646\pm 0.027$ and $0.611\pm 0.021$.

Studying the influence of noise, i.e., decoherence, on the entanglements is of importance. A wave-packet shifter, which is in practice a parallel-sided plate, was employed for this study: the shift of the wave packets was accurately controlled so that we could control the amount of the interference effect in the IFM. The situation where all genuine multipartite entanglement is lost with retaining bipartite entanglement in the system was accomplished and investigated. The analysis was achieved via inequalities $I_{k-sep}[\rho]\leq 0$ [explicitly written in Eq.~(\ref{k-sep})], which are satisfied for any $k$-separable state derived in Ref.~\cite{crit7}. Thus these inequalities enable us to reveal more refined structures. For example, if one finds a violation of the inequality $I_{3-sep}[\rho]$ for a tripartite state $\rho$, then the state is not fully separable, namely, entangled. However, it may still be ``only'' bipartite entangled, while a violation of $I_{2-sep}$ shows that the state is genuine multipartite entangled (while the criteria $I_{GHZ}$ and $I_W$ are optimized to detect the different types of genuine multipartite entanglement). The results are listed in Tab.~\ref{TableII}. The measured values are greater than $0$ and thus the non full-separability is proven, even in the noisy case. The measured values are greater than $0$ and thus the non full-separability is proven, even in the noisy case. (The fidelities also prove inseparability in these systems, albeit with a smaller experimental significance).

\begin{center}
\begin{table}[t]
    \begin{tabular}{ | l | l | l |}
    \hline
    State & Noiseless & With noise  \\ \hline
    $|GHZ\rangle$ & $I_{GHZ}=0.49\pm0.01$ & NA \\ \hline
    $|W_{sym}\rangle$ &  $I_{W}=0.47\pm0.03$ &  $I_{W}=-0.04\pm0.02$  \\ \hline
    $|W_{asym}\rangle$ &  $I_{W}=0.46\pm0.02$ &  $I_{W}=-0.01\pm0.01$  \\
    \hline
        \end{tabular}
        \caption{\label{TableI}The almost maximal violation clearly indicates multipartite entanglement and thus demonstrates the high amount of control in neutron state preparation. In the presence of noise (degrading of coherence) no multipartite entanglement is detectable.}
\end{table}
\end{center}
\vspace{-1.0cm}

\begin{center}
\begin{table}[t]
    \begin{tabular}{ | l | l | l |}
    \hline
    State & Noiseless & With noise  \\ \hline
    $|GHZ\rangle$ & $I_{3-sep}=0.49\pm0.01$ & NA \\ \hline
    $|W_{sym}\rangle$ &  $I_{3-sep}=0.65\pm0.02$ &  $I_{3-sep}=0.31\pm0.01$  \\ \hline
    $|W_{asym}\rangle$ &  $I_{3-sep}=0.45\pm0.01$ &  $I_{3-sep}=0.11\pm0.01$  \\
    \hline
        \end{tabular}
        \caption{\label{TableII}Despite the strong noise and (potential) loss of genuine multipartite entanglement the data show that the states are not fully separable. This can in this specific case also be seen by looking at the fidelity witness.}
\end{table}
\end{center}
\vspace{-1.0cm}

In the present neutron interferometer experiment specific tripartite-entangled states have been successfully generated with high fidelity and their entanglement content quantified using nonlinear witnesses. This gives a basis for studies on the mixture and transition between these states and the totally decohered state, by which characteristic properties of tripartite entanglement can be demonstrated. In such studies, influences of various kinds of decoherence on entanglement will be investigated: in particular, a method involving controllable decoherence (e.g.~like in experiment \cite{Sulyok}) will play an important role.

\section{Conclusions}

We have succeeded in generating tripartite entangled W-like state families and a GHZ-like state with high fidelity within neutron interferometry. Genuine tripartite entanglement of these states is analyzed by proper non-linear witnesses. It is demonstrated that, only by measuring few matrix elements, different types of multipartite entanglement can be proven to be present in a single-neutron system. Moreover, via the control of the system under investigation we could decohere some components of the system and thus demonstrate the loss of multipartite entanglement while inseparability is still kept. These results show that quantum correlations can indeed manifest in different subsystems of a single physical entity. Our revealed substructure of an entangled single-neutron system and high controllability opens the possibility for testing more fundamental and sophisticated properties of neutrons via exploring just this substructure, e.g. its relation to topological phases \cite{Sponar2010,Filipp2009,Bertlmann2004} or to the effect of a gravitational field \cite{topphgra}.

\section{Acknowledgment}

This work was supported by the Austrian Research Fund (FWF) and the SoMoPro programme.
DE and YH acknowledge the support from the project FWF-P21193-N20 and the Institute Laue Langevin.
MH acknowledges the support from the project FWF-P21947-N16, the EC-project IP "Q-Essence", the ERC Advanced Grant "IRQUAT" and the MC grant "Quacocos".
Research of BCH leading to these results has received a financial contribution from the European Community within the Seventh Framework Programme (FP/2007-2013) under Grant Agreement No. 229603. The research is also co-financed by the South Moravian Region.


\begin{thebibliography}{1}
\bibitem{NC} M. A. Nielsen and I. Chuang, {\it Quantum Computation and Quantum Information} (Cambridge University Press, Cambridge, 2000).

\bibitem{Hiesmayr1}  B. C. Hiesmayr {\it et al.}, Eur. Phys. J. C 72, 1856 (2012) and references therein.
\bibitem{HBB} M. Hillery, V. Buzek, and A. Berthiaume, Phys. Rev. A {\bf 59}, 1829 (1999).
\bibitem{gisin-crypt} N. Gisin, G. Ribordy, and H. Zbinden, Rev. Mod. Phys. {\bf 74}, 145 (2002).
\bibitem{SHH1} S. Schauer, M. Huber, and B. C. Hiesmayr,  Phys. Rev. A {\bf 82}, 062311 (2010).
\bibitem{qc} R. Raussendorf and H.-J. Briegel, Phys. Rev. Lett. {\bf 86}, 5188 (2001).

\bibitem{phase} S. Sachdev, {\it Quantum Phase Transitions} (Cambridge University Press, Camebridge, England, 1999).
\bibitem{helium} D. Akoury {\it et al.}, Science 9 Vol. {\bf 318}. no. 5852, p. 949 - 952, (2007).
\bibitem{hiesmayrnarnhofer} B. C. Hiesmayr, M. Koniorczyk, and H. Narnhofer, Phys. Rev. A {\bf 73}, 032310 (2006).
\bibitem{illuminati} S. M. Giampaolo, G. Adesso and F. Illuminati, Phys. Rev. Lett. {\bf 104}, 207202 (2010).

\bibitem{bio3} E. M. Gauger, E. Rieper, J. J. L. Morton, S. C. Benjamin, and V. Vedral, Phys. Rev. Lett. {\bf 106}, 040503 (2011) and references therein.

\bibitem{contextuality} S. Kochen and E. P. Specker, J. Math. Mech. {\bf 17}, 59-87 (1967); A. Peres, Phys. Lett. A {\bf 151}, 107-108 (1990); N. D. Mermin, Phys. Rev. Lett. {\bf 65}, 3373-3376 (1990); N. D. Mermin, Rev. Mod. Phys. {\bf 65}, 803-815 (1993).

\bibitem{Photon1} R. Prevedel {\it et. al.}, Phys. Rev. Lett. {\bf 103}, 020503 (2009); W. Wieczorek {\it et. al.}, Phys. Rev. Lett. {\bf 103}, 020504 (2009) and references therein.

\bibitem{ions} H. H\"affner {\it et. al.}, Nature {\bf 438}, 643 (2005) and references therein.
\bibitem{optom} S. Machnes {\it et. al.}, Phys. Rev. Lett. {\bf 108}, 153601 (2012) and references therein.

\bibitem{Martini} J. T. Barreiro, N. K. Langford, N. A. Peters, and P. G. Kwiat, Phys. Rev. Lett. {\bf 95}, 260501 (2005) and references therein.
\bibitem{Rauchbook} H. Rauch and S. A. Werner, {\it Neutron Interferometry} (Clarendon, Oxford, 2000).
\bibitem{HaseN} Y. Hasegawa {\it et. al.}, Nature {\bf 425}, 45 (2003).
\bibitem{HasePRA} Y. Hasegawa {\it et. al.}, Phy Rev. A {\bf 81}, 032121 (2010).
\bibitem{HaseNJP} Y. Hasegawa {\it et. al.}, N. J. Phys. {\bf 14}, 023039 (2012).
\bibitem{Sponar} S. Sponar {\it et. al.}, N. J. Phys. {\bf 14}, 053032 (2012).

\bibitem{crit1} M. Horodecki, P. Horodecki, and R. Horodecki, Phys. Lett. A {\bf 283}, 1 (2001); A. Acin, D. Bru{\ss}, M. Lewenstein, and A. Sanpera, Phys. Rev. Lett. {\bf 87}, 040401 (2001); P. Wocjan and M. Horodecki, Open Syst. Inf. Dyn. {\bf 12}, 331 (2005); M. Seevinck and J. Uffink, Phys. Rev. A {\bf 78}, 032101 (2008); B. C. Hiesmayr, M. Huber, and Ph. Krammer, Phys. Rev. A. {\bf 79}, 062308 (2009).

\bibitem{crit6} M. Huber, F. Mintert, A. Gabriel, and B. C. Hiesmayr, Phys. Rev. Lett. {\bf 104}, 210501 (2010).

\bibitem{crit7} A. Gabriel, B. C. Hiesmayr, and M. Huber, Quantum Inform. Comput. {\bf 10}, 0829-0836 (2010).

\bibitem{crit8} O. G\"uhne and M. Seevinck, New J. Phys. {\bf 12}, 053002 (2010).

\bibitem{GHZstate} D. M. Greenberger, M. A. Horne, and A. Zeilinger, in {\it Bell's Theorem, Quantum Theory and Conceptions of the Universe}, edited by M. Kafatos (Kluwer Academics, Dordrecht, The Netherlands, 1989), pp. 73-76.
\bibitem{Wstate} W. D\"ur, G. Vidal, and J. I. Cirac, Phys. Rev. A {\bf 62}, 062314 (2000).

\bibitem{crazychin} Z. Ma {\it et. al.}, Phys. Rev. A {\bf 83}, 062325 (2011).
\bibitem{wuetal} J. Y. Wu, H. Kampermann, D. Bru{\ss}, C. Kl\"{o}ckl, and M. Huber, Phys. Rev. A {\bf 86}, 022319 (2012).
\bibitem{KROUPA} G. Kroupa, G. Bruckner, O. Bolik, M. Zawisky, M. Hainbuchner, G. Badurek, R. J. Buchelt, A. Schricker, and H. Rauch, Nucl. Instr. and Meth. A {\bf 440}, 604 (2000).
\bibitem{Sponar08a} S. Sponar, J. Klepp, R. Loidl, S. Filipp, G. Badurek, Y. Hasegawa, and H. Rauch, Phys. Rev. A {\bf 78}, 061604(R) (2008).

\bibitem{Sulyok} G. Sulyok,  Y. Hasegawa, J. Klepp, H. Lemmel, and H. Rauch, Phys. Rev. A {\bf 81}, 053609 (2010).

\bibitem{Sponar2010} S. Sponar {\it et. al.}, Phys. Rev. A {\bf 81}, 042113 (2010).
\bibitem{Filipp2009} S. Filipp {\it et. al.}, Phys. Rev. Lett. {\bf 102}, 030404 (2009).
\bibitem{Bertlmann2004} R. A. Bertlmann, K. Durstberger, Y. Hasegawa, and B. C. Hiesmayr, Phys. Rev. A {\bf 69}, 032112 (2004).

\bibitem{topphgra} M. Zych, F. Costa, I. Pikovski, C. Brukner, Nat. Commun. {\bf 2}:505 (2011).

\end{thebibliography}
\end{document}